\documentclass[preprint,12pt, a4paper]{elsarticle}

\usepackage{amssymb}
\usepackage{hyperref}
\usepackage{braket}
\usepackage{listings}
\journal{SoftwareX}

\begin{document}
\renewcommand{\labelenumii}{\arabic{enumi}.\arabic{enumii}}

\begin{frontmatter}

 


\title{QKDNetSim+: Improvement of the Quantum Network Simulator for NS-3}

\author[udc]{David Soler\texorpdfstring{\corref{cor1}}{*}}
\ead{david.soler@udc.es}
\cortext[cor1]{Corresponding author}

\author[udc]{Iván Cillero}
\author[udc]{Carlos Dafonte}
\author[uvigo]{Manuel Fernández-Veiga}
\author[uvigo]{Ana Fernández Vilas}
\author[udc]{Francisco J. Nóvoa}

\address[udc]{University of A Coruña, Galicia, Spain}
\address[uvigo]{University of Vigo, Galicia, Spain}

\begin{abstract}
The first Quantum Key Distribution (QKD) networks are currently being deployed, but the implementation cost is still prohibitive for most researchers. As such, there is a need for realistic QKD network simulators. The  \textit{QKDNetSim} module for the network simulator NS-3 focuses on the representation of packets and the management of key material in a QKD network at the application layer. Although QKDNetSim's representation of a QKD network is insightful, some its components lack the depth that would allow the simulator to faithfully represent the behaviour of a real quantum network. In this work, we analyse QKDNetSim's architecture to identify its limitations, and we present an enhanced version of QKDNetSim in which some of its components have been modified to provide a more realistic simulation environment.
\end{abstract}


\end{frontmatter}


\section*{Metadata}

\begin{table}[!ht]
\begin{tabular}{|l|p{6.5cm}|p{6.5cm}|}
\hline
\textbf{Nr.} & \textbf{Code metadata description} & \textbf{Value} \\
\hline
C1 & Current code version & 1.0 \\
\hline
C2 & Permanent link to code/repository used for this code version & \url{https://github.com/SDABIS/qkdnetsim-dev} \\
\hline
C3  & Permanent link to Reproducible Capsule & -\\
\hline
C4 & Legal Code License   & GNU General Public License (GPL)\\
\hline
C5 & Code versioning system used & git \\
\hline
C6 & Software code languages, tools, and services used & C++, Python, NS-3 \\
\hline
C7 & Compilation requirements, operating environments \& dependencies & See GitHub repository \\
\hline
C8 & If available Link to developer documentation/manual & See GitHub Repository \\
\hline
C9 & Support email for questions & david.soler@udc.es \\
\hline
\end{tabular}
\caption{Code metadata}
\label{codeMetadata} 
\end{table}


\section{Motivation and significance}
\label{sec:introduction}

Quantum technologies have experienced a significant advance in recent years \cite{Qinternet1, Qinternet2, Qinternet3}. In specific, Quantum Key Distribution (QKD) protocols allow two nodes to agree on a key through a quantum channel, in such a way that it would be impossible for an eavesdropper to obtain the key without being detected \cite{alleaume2014using}. This key can then be used to encrypt communications between the two nodes.

The number of implemented QKD networks is currently very small due to the high cost of the required material and the lack of maturity in the technology. Thus, researchers must employ simulators that imitate the behaviour of a quantum network. There are multiple alternatives depending on the scope of the research \cite{qkd-sims}: some simulators focus on representing the physical layer of the quantum channel, while others allow users to define entire networks in which nodes can execute QKD between them.

The network simulator NS-3 is widely used in the scientific and educational communities due to its level of detail and its customising capabilities. There exists a module implemented for NS-3 for the simulation of quantum networks, named QKDNetSim \cite{qkdnetsim-original}. The advantages of QKDNetSim over other simulators come from the granularity of NS-3: this simulator allows for in-depth configuration of every component, and packets sent over the simulated network are fully defined, including headers for all protocols involved. Unlike other quantum network simulators, QKDNetSim focuses on a network perspective and the managing of the key material generated through QKD. The layered structure of NS-3 also allows QKDNetSim to represent QKD networks at the application level, in which multiple users and programs share common QKD resources such as quantum channels and key buffers. This level of detail, when applied to QKDNetSim's simulation of quantum networks, could help newcomers to understand the fundamentals of quantum communications. 

However, QKDNetsim contains some limitations that affect its ability to faithfully represent the behavior of a real quantum network. The main shortcomings of QKDNetSim are in key management: the nodes do not adequately process the key material that they receive and the cryptography handler does not function properly. These aspects are crucial to the architecture of QKD networks, so QKDNetSim's inability to simulate them imposes a significant limitation on its usefulness.

For this work we present an enhanced version of the quantum network simulation module QKDNetSim. To this end, we will start with an analysis of QKDNetSim: of which elements it is composed and to which degree it imitates the behaviour of a real quantum network, including the limitations and errors that prevent QKDNetSim from achieving its purposed objectives. Our implementation will provide enhancements to overcome the mentioned limitations while maintaining the architecture of the simulation module. The enhanced version of QKDNetSim can be employed by researchers to study the behaviour of QKD networks, with a focus on key synchronisation and management. It can also be used as a first approach to the design of complex QKD networks and to analyse their performance.

\subsection{QKD network architecture}
\label{sec:net}

In experimental settings, the execution of QKD protocols is very slow and thus it would be very inefficient to block the generation of a packet to obtain key material through QKD \cite{qkd-network}. For that reason, it is common to asynchronously execute the QKD protocol at a previous time, and store the resulting key material in a buffer to be consumed when needed \cite{keypool}. As a result, the processes of generating and consuming key material are decoupled, at the cost of managing the synchronisation of buffers of adjacent nodes.

\begin{figure}[!htb]
    \centering
    \includegraphics[width=\columnwidth]{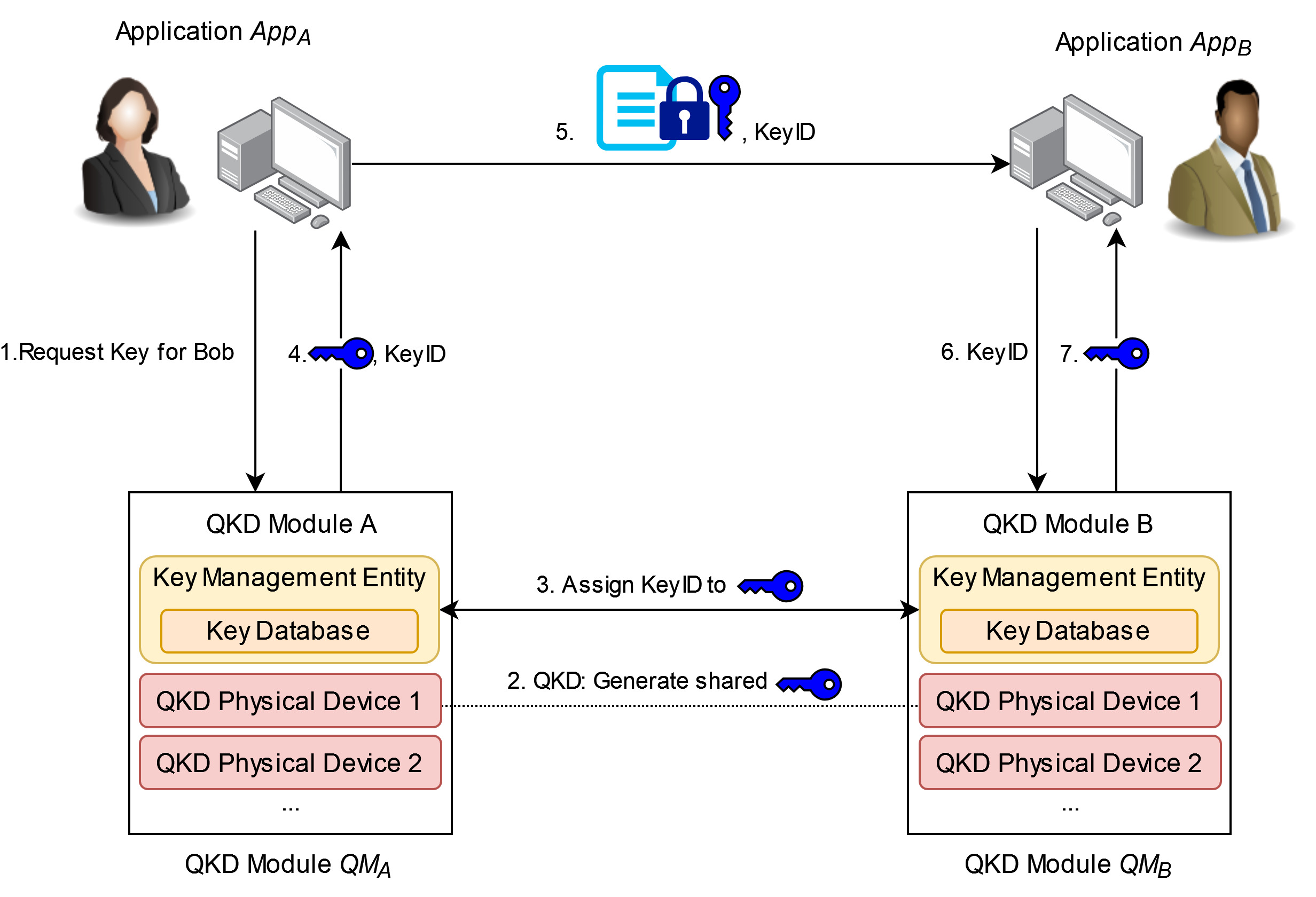}
    \caption{Diagram of a QKD communication.}
    \label{fig:etsi}
\end{figure}

Due to the high cost of deploying quantum communications infrastructure, these devices are usually shared between multiple Applications. There have been efforts to standardise the communication between Applications and QKD devices. In the ETSI GS QKD standards 004 \cite{etsi004} and 014 \cite{etsi014}, a QKD Module is composed of multiple QKD Physical Devices (possibly belonging to different manufacturers) and a Key Management Entity, or $KME$. The KME contains a Key Database and is tasked with managing the key material that is shared through the QKD Physical Devices. An example communication flow between Applications is shown in Figure \ref{fig:etsi}: their respective QKD Modules are tasked with generating the shared key material through a QKD protocol (such as BB84 \cite{bb84}) and assigning it a KeyID, which the Applications can then use to retrieve the cryptographic key.

\subsection{QKD Simulators}
\label{sec:sims}

Due to the experimental state of QKD, most researchers rely on simulators to study the behaviour and structure of quantum networks. There are a variety of available simulators \cite{qkd-sims} that focus on different aspects of quantum networks, whether it be the network structure, the representation of quantum states or the execution of QKD protocols.

\textit{Qunetsim} \cite{qunetsim} is a simulator written in Python that allows users to represent quantum networks. The nodes can exchange not only classical messages, but also qubits that are simulated through different backends such as $SimulaQron$. Qunetsim mainly focus on simulating the quantum channels that connect nodes and implementing different QKD protocols and methods of transmitting information through qubits, including the use of entangled states. Because of this low-level approach, Qunetsim does not provide functionalities for key management or buffer synchronisation between QKD nodes. The network is functional, but not completely simulated as packets do not contain protocol headers as in real networks.

\textit{NetSquid} \cite{netsquid} is another alternative written in Python. It also focuses on the representation of qubits and the simulation of quantum channels, including the possibility of adding delay, noise and loss to imitate real conditions. Unlike Qunetsim, NetSquid does formally represent the behaviour of a real network by employing a discrete-event simulation engine. However, it also leaves the management of key material to be implemented by users.

\section{Software description}

\subsection{Original QKDNetSim's Architecture}

The module QKDNetSim adds to NS-3 the possibility of creating quantum channels between pairs of nodes. To this end, the following components are added to each pair of nodes:

\begin{figure}
    \centering
    \includegraphics[width=0.8\columnwidth]{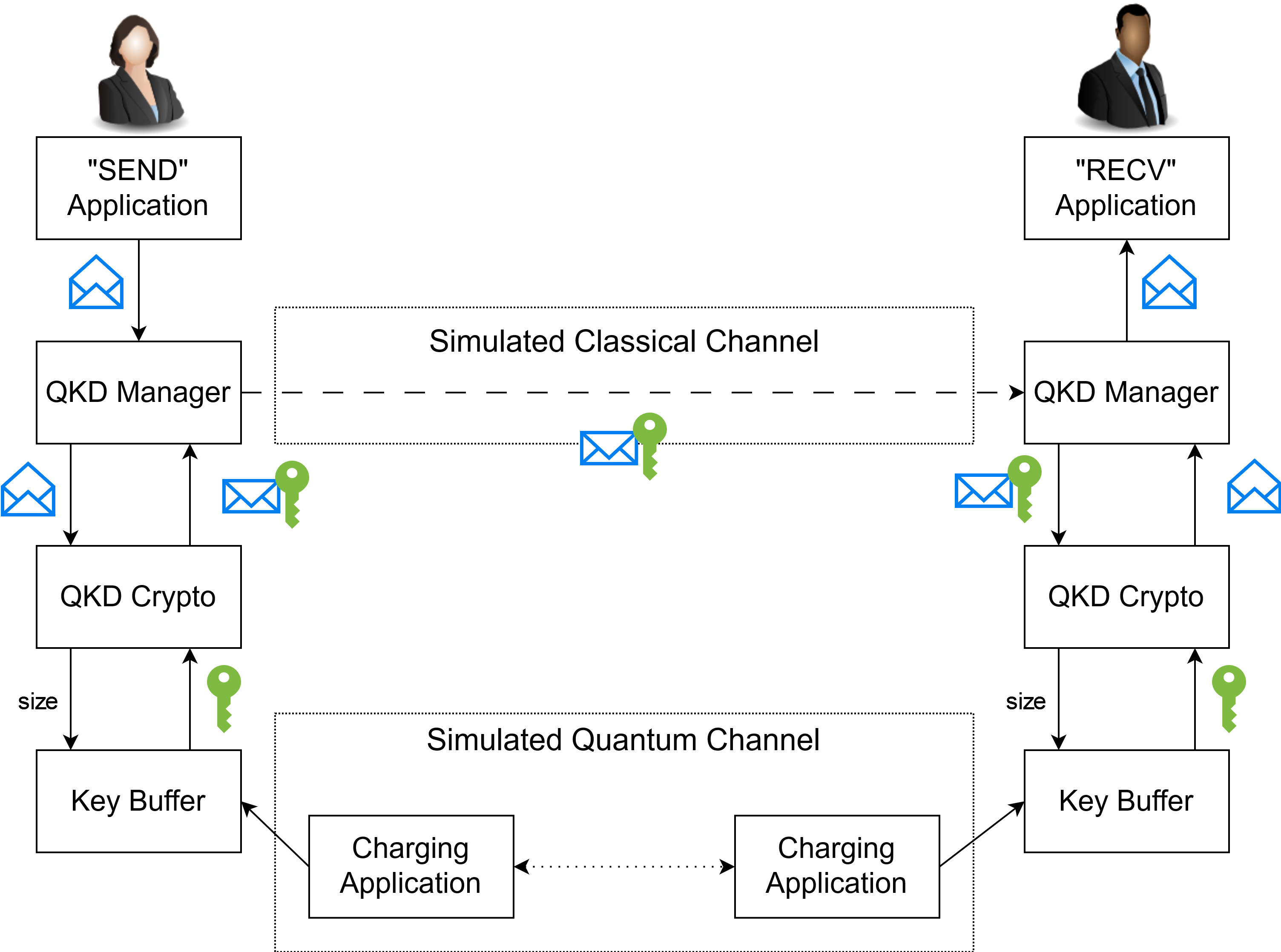}
    \caption{Architecture of QKDNetSim.}
    \label{fig:arch}
\end{figure}

\begin{itemize}
    \item \textbf{Send/Receive Applications}: they simulate programs that create packets to be encrypted by the rest of QKDNetSim's components. When they are generated or received, they are sent to the Manager to perform the pertinent cryptographic operations.
    \item \textbf{QKD Manager}: the central component of the module, which serves as connection between the others. The Manager processes incoming and outgoing packets, identifies which operations need to be performed and calls the pertinent component to execute them.
    \item \textbf{Cryptography Handler}: receives petitions from the Manager to encrypt or decrypt packets. It implements multiple cryptographic algorithms, and has access to the Key Buffer. Corresponds with the Object \textit{QKDCrypto}
    \item \textbf{Simulated Quantum Channel}: imitates the behaviour of a quantum channel. Each of the nodes possesses a \textit{Charging Application}, which constantly generate new shared key material.
    \item \textbf{Key Buffer}: Stores key material transmitted through the quantum channel for future use.
\end{itemize}

The interaction between elements is shown in Figure \ref{fig:arch}. 

Unlike the simulators mentioned in Section \ref{sec:sims}, QKDNetSim focuses on the structure of QKD nodes and the components that manage the key material that is generated through QKD protocols. The structure of QKD nodes, with the inclusion of Key Buffers as a fundamental component, is compatible with ETSI's definition of a QKD Module, as represented in Figure \ref{fig:etsi}.

\begin{figure}
    \centering
    \includegraphics[width=0.9\columnwidth]{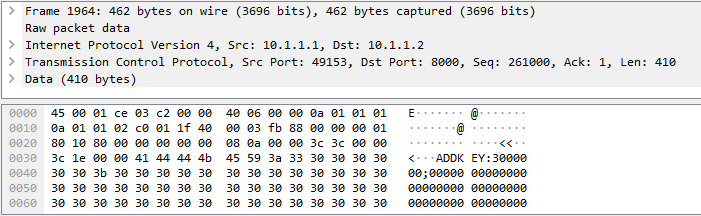}
    \caption{Packet containing key material sent by Charging Application.}
    \label{fig:addkey}
\end{figure}

As mentioned, the quantum channel is simulated through the \textit{Charging Applications} that are installed for each pair of connected nodes. Figure \ref{fig:addkey} shows the contents of a packet exchanged between the Charging Applications of two adjacent nodes, which includes (after the label $ADDKEY$) the key material that will be added to the buffers.

QKDNetSim uses buffers to store key material, as introduced in Section \ref{sec:net}. the Charging Applications are constantly generating new key material and storing it into their respective Key Buffers. Whenever the Send Application generates a new packet, some of the previously generated key material is consumed to encrypt it. 

\begin{figure}
    \centering
    \includegraphics[width=\columnwidth]{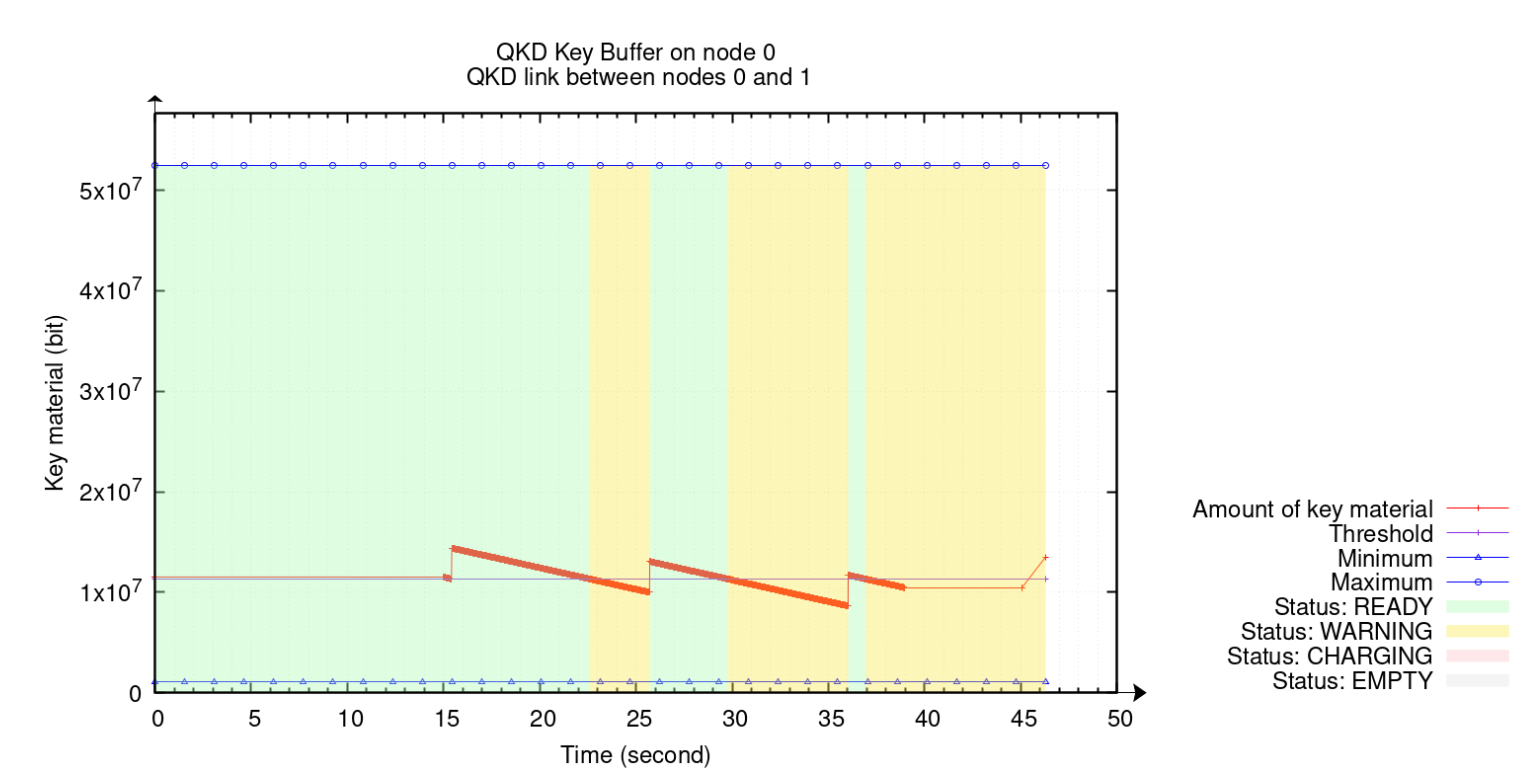}
    \caption{Amount of available key material inside the Key Buffer throughout the simulation.}
    \label{fig:buffer}
\end{figure}

Figure \ref{fig:buffer} shows the amount of key material stored inside a Key Buffer during a simulation. It increases periodically when the Charging Applications creates new key material, and it decreases when new packets are encrypted. It is possible to manually define the maximum and minimum amounts of key material that the Key Buffer can hold, as well as the initial amount. As shown in Figure \ref{fig:buffer}, the Simulated Quantum Channel only generates new key material when the current amount is below a "Threshold" value.

\subsection{QKDNetSim's Shortcomings}

QKDNetSim correctly identifies the components that take part in a QKD network. However, the implementation of these components lacks realism in some aspects, which prevent QKDNetSim from providing a faithful simulation. We highlight the following shortcomings of QKDNetSim:

\paragraph{\textbf{The messages sent through the Simulated Quantum Channel are all identical.}}

The Simulated Quantum Channel is tasked with the creation of new key material between two nodes. However, all messages exchanged between the Charging Applications of adjacent nodes contain only a string of '0's, as can be seen in Figure \ref{fig:addkey}. Since all the key material that is created is the same, all packets are encrypted with the same key. The receiver's Charging Application does not process $ADDKEY$ messages, since it already knows that the key is a string of '0's. This defeats the purpose of QKD and makes QKDNetSim a less realistic simulator. 

\paragraph{\textbf{The Key Buffers do not store keys.}}

The key material generated through the Simulated Quantum Channel is not stored anywhere. This should be the function of the Key Buffer: instead, it only stores a number which represents the \textit{amount} of key material it should have. 

Whenever the Simulated Quantum Channel tries to insert new key material in the Key Buffer, it is discarded and the \textit{amount} is increased. A similar process occurs when the Manager requests key material from the Buffer: a new string of '0's is returned and the \textit{amount} is decreased by its length. The plot shown in Figure \ref{fig:buffer} represents this \textit{amount}, since it is the only metric the Buffer can provide.

Since there is no real key material inside the Buffers, there is also no mechanism for ensuring, maintaining or recovering synchronicity between Key Buffers of connected nodes. This makes QKDNetSim less realistic, as it does not represent a problem that needs to be addressed in real quantum networks.

\paragraph{\textbf{Encryption is disabled and does not work.}}

The Cryptography Handler provides implementations for the encryption algorithms AES and One-Time Pad (OTP). However, these implementations contain errors: they incorrectly assume keys are represented as an array of bytes, when they are an array of bits. Thus, whenever the simulator tries to encrypt a packet, it crashes.

Even if these errors are corrected and packets are adequately encrypted, the PCAP still shows them in plaintext. The fact that contents of the PCAP do not correspond to the real messages that were exchanged makes the simulation harder to understand.

\subsection{Description of enhancements to QKDNetSim}
\label{sec:impl}

As we have shown, QKDNetSim is limited in its representation of QKD networks since it does not adequately represent tasks related to key management, which are fundamental operations in real scenarios. In our improved version of QKDNetSim we provide the following enhancements to the aforementioned shortcomings:

\paragraph{\textbf{Key Generation and Transmission}}

The Charging Applications now adequately process incoming $ADDKEY$ messages. As in the original module, one of the nodes is designated as "Primary" and the other as "Secondary": the Primary node is tasked with generating the key material and sending the $ADDKEY$ messages, while the Secondary receives and processes them.

As mentioned, in the original QKDNetSim all keys are generated by the Charging Applications as a string of '0's. In our implementation we provide a new component named "QKD Random Generator" for the generation of key material. We use NS-3's PRNG as the default source of randomness, but we also provide the option of using a real Quantum Random Number Generator (QRNG)~\cite{qrng2, qrng3, qrng4}, which are commonly used as a source of entropy in quantum networks.

We have employed IDQuantique's \textit{Quantis QRNG}, which is accessible through a USB interface and has a key generation rate of about 4 Mb/s. Our QKD Random Generator component performs calls to Quantis' C++ API to generate the key material that is to be transmitted through the Simulated Quantum Channel.

\paragraph{\textbf{Key Storage}}

The Key Buffers are now composed of two different structures:

\begin{enumerate}
    \item \textit{Raw key material storage}: Key material generated by the Simulated Quantum Channel that has not been assigned a KeyID yet. Its capacity is regulated by the "MAX", "MIN" and "THRESHOLD" parameters shown in Figure \ref{fig:buffer}
    \item \textit{Key Database}: Set of indexed keys with information about their KeyID and length. 
\end{enumerate}

The process of obtaining key material from the raw storage and inserting it into the Key Database with a KeyID is called \textit{reserving}. In our implementation, this is performed by the Cryptography Handler whenever it requires a key of any length for either encryption or authentication of Application packets. The reservation of key material is done in the sender and receiver at the same time, so that both nodes can access the same key with a specific KeyID. The keys are used only once and then are removed from the Key Database.

\paragraph{\textbf{Encryption}}

Whenever a packet generated by an Application reaches the Cryptography Handler, it is encrypted and authenticated, including any TCP/IP headers that it contains. Then, a \textit{QKDHeader} is added, which includes the length of the packet, the MessageID, and the KeyIDs for the encryption and authentication keys. The QKDHeader was also present in the original QKDNetSim, but since the Key Buffers did not contain a Key Database, the KeyIDs had no meaning. Therefore, the values of the QKDHeader were mostly ignored.

\begin{figure}[!htb]
    \centering
    \includegraphics[width=\columnwidth]{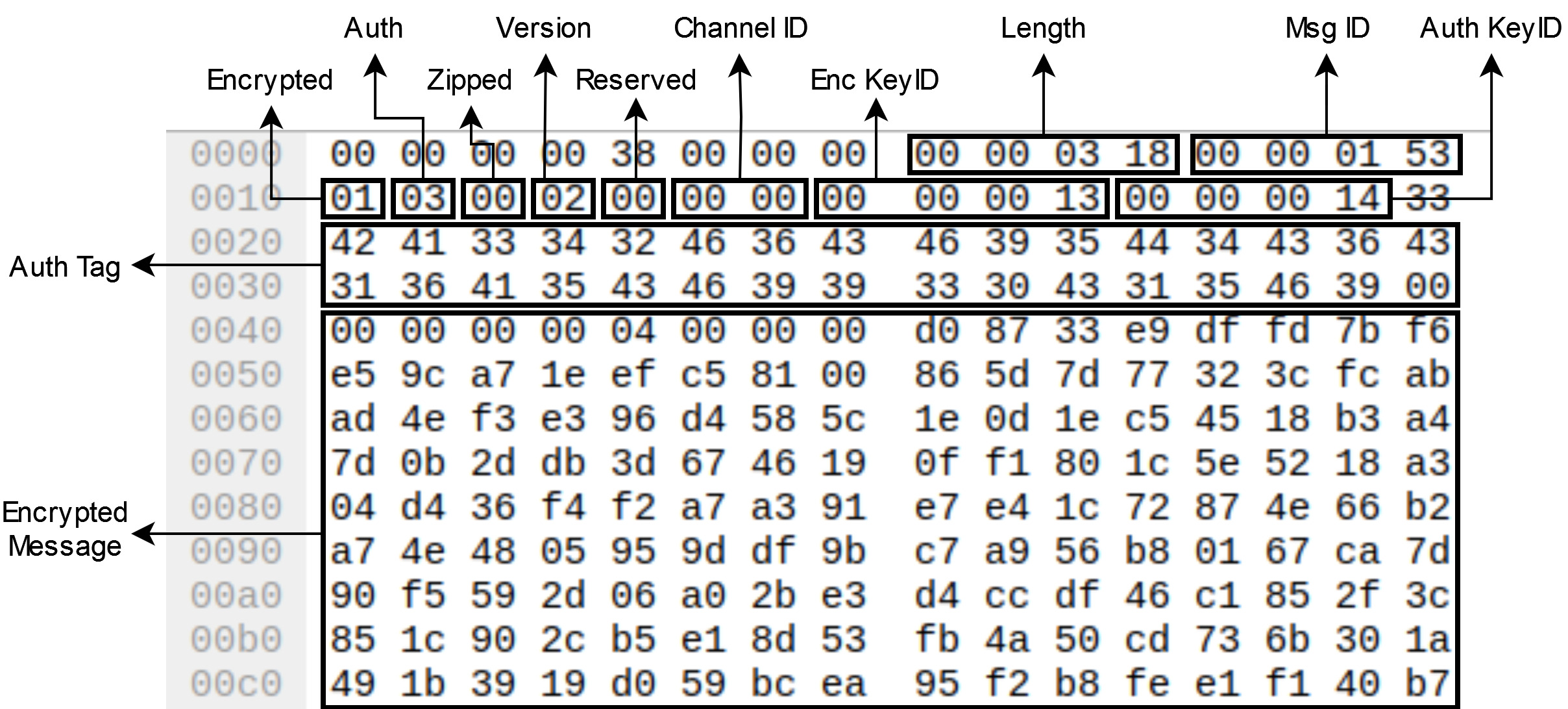}
    \caption{Contents of an encrypted message sent by the Send Application.}
    \label{fig:qkdheader}
\end{figure}

The original QKDNetSim omits the logging of encrypted messages into a PCAP file, as it only includes their unencrypted version in packet captures. Of particular interest are the values inside the QKDHeader. Figure \ref{fig:qkdheader} shows an example packet as logged by the simulator. The values for "Encrypted" and "Auth" represent the employed algorithms, which for this packet are OTP and VMAC, respectively.

\section{Illustrative example}

We have prepared an example execution to demonstrate the behaviour of the enhanced QKDNetSim module. The simulation consists of two different nodes, with IPs in the 10.1.1.0/24 network: a sender and a receiver. The sender will generate 5 packets and send them to the receiver, encrypted with key material generated through the Simulated Quantum Channel. Listing \ref{lst:output} shows the output of an example execution of QKDNetSim.

\begin{lstlisting}[caption={Output of text execution.}, label={lst:output}, captionpos=b]
Source IP address: 10.1.1.1 
Destination IP address: 10.1.1.2

+0.000000000s -1 QKDRandomGenerator:printCardsInfo(0x5573d19b9c80, "Displaying cards info:")
+0.000000000s -1 QKDRandomGenerator:printCardsInfo(0x5573d19b9c80, "* Searching for USB devices...")
+0.000000000s -1 QKDRandomGenerator:_printCardsInfo(0x5573d19b9c80, "  Found ", 1, " card(s)")
+0.000000000s -1 QKDRandomGenerator:_printCardsInfo(0x5573d19b9c80, "  - Details for device #", 0)
+0.000000000s -1 QKDRandomGenerator:_printCardsInfo(0x5573d19b9c80, "      driver version: ", 0, ".", 1)
+0.000000000s -1 QKDRandomGenerator:_printCardsInfo(0x5573d19b9c80, "      core version: ", , 60b1c01)
+0.000000000s -1 QKDRandomGenerator:_printCardsInfo(0x5573d19b9c80, "      serial number: ", , "206361A410")
+0.000000000s -1 QKDRandomGenerator:_printCardsInfo(0x5573d19b9c80, "      manufacturer: ", , "id Quantique")
+0.000000000s -1 QKDRandomGenerator:_printCardsInfo(0x5573d19b9c80, "      module ", 0, ": ", "found", " ", "(enabled)")
QKDCrypto:QKDCrypto(0x558a0f5a4d60) 
+0.000000000s -1 QKDBuffer:QKDBuffer(0x558a0f5a8460, 0, 1) 
+0.000000000s -1 QKDBuffer:Init(0x558a0f5a8460) 
+0.000000000s -1 QKDBuffer:QKDBuffer(0x558a0f5a8770, 0, 1) 
+0.000000000s -1 QKDBuffer:Init(0x558a0f5a8770)
+0.000000000s -1 QKDRandomGenerator:generateStream(0x7fff0becb050, "Requesting ", 51000, " bytes: ")
+0.000000000s -1 QKDRandomGenerator:generateStream(0x7fff0becb050, "Single call")
+0.000000000s -1 QKDBuffer:AddKeyMaterial(0x558a0f5a8460, "m_Mcurrent:", 0, "size:", 51000, "key material[0-30]:", 82, 228, 78, 97, 249, 182, 105, 0, [...], "key material[(end - 30) - end]:", [...], 149, 189, 196, 234, 197, 154, 69, 235) 
+0.000000000s -1 QKDBuffer:AddKeyMaterial(0x558a0f5a8460, "m_Mcurrent:", 51000, "buffer final material:", [...], 149, 189, 196, 234, 197, 154, 69, 235) 
+0.000000000s -1 QKDBuffer:AddKeyMaterial(0x558a0f5a8460, " Adding new Key Material: ", 51000, " bytes") 
+0.000000000s -1 QKDBuffer:AddKeyMaterial(0x558a0f5a8770, "m_Mcurrent:", 0, "size:", 51000, "key material[0-30]:", 82, 228, 78, 97, 249, 182, 105, 0, [...], "key material[(end - 30) - end]:", [...], 149, 189, 196, 234, 197, 154, 69, 235) 
+0.000000000s -1 QKDBuffer:AddKeyMaterial(0x558a0f5a8770, "m_Mcurrent:", 51000, "buffer final material:", [...], 149, 189, 196, 234, 197, 154, 69, 235) 
+0.000000000s -1 QKDBuffer:AddKeyMaterial(0x558a0f5a8770, " Adding new Key Material: ", 51000, " bytes") 
+0.000000000s -1 QKDBuffer:InitTotalGraph(0x558a0f5a8460) 
+0.000000000s -1 QKDBuffer:InitTotalGraph(0x558a0f5a8770) 
 
[...] 
 
QKDCrypto:ProcessOutgoingPacket(0x558a0f5a4d60, "***** ENCRYPTION MODE *****", 1) 
QKDCrypto:ProcessOutgoingPacket(0x558a0f5a4d60, "SrcBuffer") 
+20.256000000s 0 QKDBuffer:ReserveKeyMaterial(0x558a0f5a8460, " Reserved Key with ID ", 19, "of size", 720) 
+20.256000000s 0 QKDBuffer:FetchKeyByID(0x558a0f5a8460, "Fetching Key with ID: ", 19, "	Value: ", "D08733E9CBFD7BF6[...]") 
QKDCrypto:ProcessOutgoingPacket(0x558a0f5a4d60, "DstBuffer") 
+20.256000000s 0 QKDBuffer:ReserveKeyMaterial(0x558a0f5a8770, " Reserved Key with ID ", 19, "of size", 720) 
+20.256000000s 0 QKDBuffer:FetchKeyByID(0x558a0f5a8770, "Fetching Key with ID: ", 19, "	Value: ", "D08733E9CBFD7BF6[...]") 
+20.256000000s 0 QKDBuffer:DeleteKeyID(0x558a0f5a8460, "KeyID is deleted from 'm_keys'", 19) 
QKDCrypto:ProcessOutgoingPacket(0x558a0f5a4d60, "Encryption completed!") 
QKDCrypto:ProcessOutgoingPacket(0x558a0f5a4d60, "***** AUTHENTICATION MODE *****", 3) 
QKDCrypto:ProcessOutgoingPacket(0x558a0f5a4d60, "SrcBuffer") 
+20.256000000s 0 QKDBuffer:ReserveKeyMaterial(0x558a0f5a8460, " Reserved Key with ID ", 20, "of size", 32) 
+20.256000000s 0 QKDBuffer:FetchKeyByID(0x558a0f5a8460, "Fetching Key with ID: ", 20, "	Value: ", "71B864E55B1D5327[...]") 
QKDCrypto:ProcessOutgoingPacket(0x558a0f5a4d60, "DstBuffer") 
+20.256000000s 0 QKDBuffer:ReserveKeyMaterial(0x558a0f5a8770, " Reserved Key with ID ", 20, "of size", 32) 
+20.256000000s 0 QKDBuffer:FetchKeyByID(0x558a0f5a8770, "Fetching Key with ID: ", 20, "	Value: ", "71B864E55B1D5327[...]") 
QKDCrypto:ProcessOutgoingPacket(0x558a0f5a4d60, "Adding AUTHTAG to the packet!", "3BA342F6CF95D4C6C16A5CF9930C15F9", 32) 
+20.256000000s 0 QKDBuffer:DeleteKeyID(0x558a0f5a8460, "KeyID is deleted from 'm_keys'", 20) 
QKDCrypto:ProcessOutgoingPacket(0x558a0f5a4d60, "Final outgoing packet from QCrypto:", "PacketID:", 340, "of size", 792, "MessageID:", 339, "QKDHeaderLength:", 792, "Encryption KeyID:", 19, "Auth KeyID: ", 20) 
0x558a0f5a4d60	QKDBuffer: 	0	TOSband: 	1	KeySize: 	0 
 
QKDCrypto:ProcessIncomingPacket(0x558a0f5a4d60, "PacketID:", 340, "PacketSize:", 792, 0x558a0f5a8770, "ChannelID:", 0) 
+20.259270400s 1 QKDBuffer:FetchKeyByID(0x558a0f5a8770, "Fetching Key with ID: ", 20, "	Value: ", "71B864E55B1D5327[...]") 
+20.259270400s 1 QKDBuffer:DeleteKeyID(0x558a0f5a8770, "KeyID is deleted from 'm_keys'", 20) 
+20.259270400s 1 QKDBuffer:FetchKeyByID(0x558a0f5a8770, "Fetching Key with ID: ", 19, "	Value: ", "D08733E9CBFD7BF6[...]") 
+20.259270400s 1 QKDBuffer:DeleteKeyID(0x558a0f5a8770, "KeyID is deleted from 'm_keys'", 19) 
QKDCrypto:ProcessIncomingPacket(0x558a0f5a4d60, "Decrytion completed!", 0x558a0f6b10a0) 
 
[...]

+20.236610455s 0 QKDChargingApplication:PrepareOutput(0x5573d19b9820, +20236610455.0ns, ADDKEY3072000)
+20.236610455s 0 QKDRandomGenerator:generateStream(0x5573d19b9c80, "Requesting ", 500, " bytes: ")
+20.236610455s 0 QKDRandomGenerator:generateStream(0x5573d19b9c80, "Single call")
+20.236610455s 0 QKDBuffer:AddKeyMaterial(0x558a0f5a8460, "m_Mcurrent:", 50960, "size:", 500, "key material:", 220, 156, 42, 31, 103, 123, 152, 13, [...], 17, 171, 245, 227, 25, 254, 225, 103) 
+20.236610455s 0 QKDBuffer:AddKeyMaterial(0x558a0f5a8460, "m_Mcurrent:", 51460, "key_material size:", 51460) 
+20.236610455s 0 QKDBuffer:AddKeyMaterial(0x558a0f5a8460, "m_Mcurrent:", 51460, "buffer final material:", 142, 213, 170, 198, 227, 249, 56, 222, 197, 143, 58, 32, 102, 92, 195, 213, 75, 112, 124, 118, 216, 231, 17, 171, 245, 227, 25, 254, 225, 103) 
+20.236610455s 0 QKDBuffer:AddKeyMaterial(0x558a0f5a8460, " Adding new Key Material: ", 51460, " bytes") 
QKDChargingApplication:PrepareOutput(0x5573d19b9820, "The realKey was added to the SrcBuffer", 0) 
QKDChargingApplication:DataSend(0x5573d19b9820) 

QKDChargingApplication:HandleRead(0x55c03da58990, 0x55c03da66e80, "PACKETID: ", 325, " of size: ", 511) 
QKDChargingApplication:HandleRead(0x55c03da58990, "At time ", 20.246, "s packet sink received ", 511, " bytes from ", 10.1.1.1, " port ", 49153, " total Rx ", 2555, " bytes") 
QKDChargingApplication:ProcessIncomingPacket(0x55c03da58990, "Adding new key to DstBuffer") 
+20.246034086s 1 QKDBuffer:AddKeyMaterial(0x558a0f5a8770, "m_Mcurrent:", 50960, "size:", 500, "key material:", 220, 156, 42, 31, 103, 123, 152, 13, [...], 17, 171, 245, 227, 25, 254, 225, 103) 
+20.246034086s 1 QKDBuffer:AddKeyMaterial(0x558a0f5a8770, "m_Mcurrent:", 51460, "key_material size:", 51460) 
+20.246034086s 1 QKDBuffer:AddKeyMaterial(0x558a0f5a8770, "m_Mcurrent:", 51460, "buffer final material:", 142, 213, 170, 198, 227, 249, 56, 222, 197, 143, 58, 32, 102, 92, 195, 213, 75, 112, 124, 118, 216, 231, 17, 171, 245, 227, 25, 254, 225, 103) 
+20.246034086s 1 QKDBuffer:AddKeyMaterial(0x558a0f5a8770, " Adding new Key Material: ", 51460, " bytes") 
\end{lstlisting}

In the first section, the QKDBuffer instances are initialised and some key material is added to both, whose size is 5100 bytes in this example. In order to generate the key material, the QKDRandomGenerator is created. The trace shows the first and last bytes that are inserted into both buffers. 

The second section shows the processing of a packet. The sender first notifies both buffers that they must reserve two different keys for encryption and authentication, whose KeyIDs are 19 and 20, respectively. The encryption algorithm is One-Time-Pad and the authentication algorithm is VMAC, so the key sizes are chosen accordingly. The process of reserving key material consists of removing the requested amount of bytes from the Key Buffer and assigning them a KeyID. Once the Key Buffers are notified, the sender encrypts and authenticates the message (including IPv4 and TCP headers) and adds a QKDHeader which includes the KeyIDs of the employed keys. The receiver then processes this packet, labelled with PacketID 340, and obtains the required keys from its Key Buffer using the KeyIDs to complete the decryption process. The encrypted message that appears in this execution trace is also shown in Figure \ref{fig:qkdheader}, where all QKDHeader values are identified.

The last section shows an exchange between the Charging Applications to add new key material to both nodes' Key Buffers. The sender's Charging Application generates a random string of 500 bytes through the QKDRandomGenerator and adds it to its buffer. The key material is then encapsulated in an ADDKEY message and sent to the receiver's Charging Application, which processes the message and inserts the same random string in its Key Buffer. The trace shows the trailing bytes of both Key Buffers, which correspond to the key material that was inserted.

\section{Impact}

\begin{table*}
    \centering
    \resizebox{\textwidth}{!}{
    \begin{tabular}{cccccc}
    \hline
    Ref.              & Name               & Network Simulation      & QKD Headers \& packets & Key Management & Classical Encryption  \\ \hline
    \cite{qunetsim} & QuNetSim             & Ad-hoc                  & No              & No                      & No \\
    \cite{netsquid} & NetSquid             & Discrete-event          & No              & No                      & No \\
    \cite{qkdnetsim-original} & QKDNetSim  & Discrete-event          & Yes             & Yes (limited)                      & Non-functional \\
    This work & -                          & Discrete-event          & Yes             & Yes       & Yes                       \\
    \end{tabular}}
    \caption{Comparison between other QKD network simulators and our implementation.}
    \label{table:comparison}
\end{table*}
Our implementation of an enhanced QKDNetSim provides a faithful representation of QKD networks, which allows users to better understand the operations involved in them, specially those related to key management at the application layer.

Table \ref{table:comparison} highlights the difference between existing QKD network simulators, including QKDNetSim in its original state, and our implementation. As shown, QKD network simulators usually place its focus on representing a quantum channel in order to simulate QKD algorithms or entanglement of qubit states. While they are undoubtedly very useful simulators, they leave out significant components of QKD networks like key management. In contrast, in our simulator the shared key material generated through QKD protocols is put to use: we allow for different Applications to employ this cryptographic keys without needing to interact with the physical layer of QKD. This also requires the use of different structures like Key Buffers to decouple the processes of obtaining the key material and using it, and to perform synchronisation with buffers of adjacent nodes.

Our enhancements over the original QKDNetSim provide a more realistic simulation of QKD networks. We maintain QKDNetSim's overall architecture while improving each component separately. In our enhanced version of QKDNetSim, the simulator uses randomly-generated key material (with the option of employing a real QRNG) instead of dummy packets filled with zeros. The inclusion of real key material also implies that the Simulated Quantum Channel must be able to process incoming QKD packets and that Key Buffers need to be actively synchronised.

\section{Conclusions}

In this work, we have analysed NS-3's quantum network simulation module QKDNetSim, describing its main components and identifying its shortcomings and their impact in the quality of the simulation of a QKD network. The shortcomings that we analyse in this document are related to key management and encryption which, as we have identified, are the elements that mainly differentiate QKDNetSim from other quantum network simulators. Our implementation of an enhanced QKDNetSim maintains the module's overall structure while overcoming the limitations of the Key Buffer, Cryptography Handler and Simulated Quantum Channel. We also provide the option of employing a real QRNG as a source of randomness for the Simulated Quantum Channel. 

\section*{Declaration of competing interest}

The authors declare that they have no known competing financial interests or personal relationships that could have appeared to influence the work reported in this paper.

\section*{Acknowledgement}

The work is funded by the Plan Complementario de Comunicaciones Cu\'anticas, Spanish Ministry of Science and Innovation(MICINN), Plan de Recuperación NextGeneration, European Union (PRTR-C17.I1, CITIC Ref. 305.2022), and Regional Government of Galicia (Agencia Gallega de Innovación, GAIN, CITIC Ref. 306.2022) D.S. acknowledges support from Xunta de Galicia and the European Union (European Social Fund - ESF) scholarship [ED481A-2023-219].

This work is part of the project TED2021-130369B-C31 and TED2021-130492B-C21 funded by MCIN/AEI/10.13039/501100011033 and by “ERDF A way of making Europe”.

\bibliographystyle{elsarticle-num} 
\bibliography{main}

\end{document}